\documentclass[final]{cvpr}

\usepackage{graphicx}
\usepackage{footmisc}
\usepackage{booktabs}       
\usepackage{amsfonts}       
\usepackage{nicefrac}       
\usepackage{microtype}      
\usepackage{color}
\usepackage{bm}
\usepackage{amsmath}
\usepackage{wrapfig}
\usepackage{graphicx}
\usepackage{setspace}
\usepackage[super]{nth}
\usepackage{siunitx}

\usepackage{caption}
\def\cvprPaperID{1468}
\usepackage{blindtext}
\usepackage{marvosym}
\usepackage{times}
\usepackage{epsfig}
\usepackage{graphicx}
\usepackage{amsmath}
\usepackage{amssymb}
\usepackage{colortbl}
\definecolor{mygray}{gray}{.88}
\usepackage{latexsym}
\usepackage{CJKutf8}

\usepackage[pagebackref=true,breaklinks=true,colorlinks,bookmarks=false]{hyperref}
\usepackage{hyperref}
\usepackage{bm}

\usepackage{caption}
\usepackage[caption=false]{subfig}
\usepackage{multirow}
\usepackage{multicol}
\usepackage{adjustbox}
\usepackage{bm}

\def\confYear{CVPR 2022}

\def\cvprPaperID{1468} 

\begin{document}

\title{\emph{Bailando}: 3D Dance Generation by Actor-Critic GPT \\ with Choreographic Memory}

\author{\vspace{-30pt}
\\{Li Siyao$^{1}$} $\,\,\,\,\,\,\,$ Weijiang Yu$^{2}$ $\,\,\,\,\,\,\,$ Tianpei Gu$^{3, 4}$ $\,\,\,\,\,\,\,$ Chunze Lin$^{4}$ \\  Quan Wang$^{4}$ $\,\,\,\,\,\,\,\,$ Chen Qian$^{4}$
$\,\,\,\,\,\,\,\,$ Chen Change Loy$^{1}$
$\,\,\,\,\,\,\,\,$ Ziwei Liu\textsuperscript{1~\Letter}\\

$^{1}$S-Lab, Nanyang Technological University \\  $^{2}$Sun Yat-Sen University $\,\,\,\,\,\,$  $^{3}$UCLA $\,\,\,\,\,\,$ $^{4}$SenseTime Research  \\

{\tt\small siyao002@e.ntu.edu.sg} $\,\,\,$ {\tt\small weijiangyu8@gmail.com} $\,\,\,$ {\tt\small gutianpei@ucla.edu } \\ {\tt\small \{linchunze, wangquan, qianchen\}@sensetime.com}  $\,\,\,$ {\tt\small \{ccloy, ziwei.liu\}@ntu.edu.sg} 
%
}

\twocolumn[{%
\renewcommand\twocolumn[1][]{#1}%
\maketitle
\begin{center}
    \centering
    \small{
     \vspace{-23pt}
    \includegraphics[width=0.92\linewidth, trim=0pt 0pt 30pt 120pt, clip]{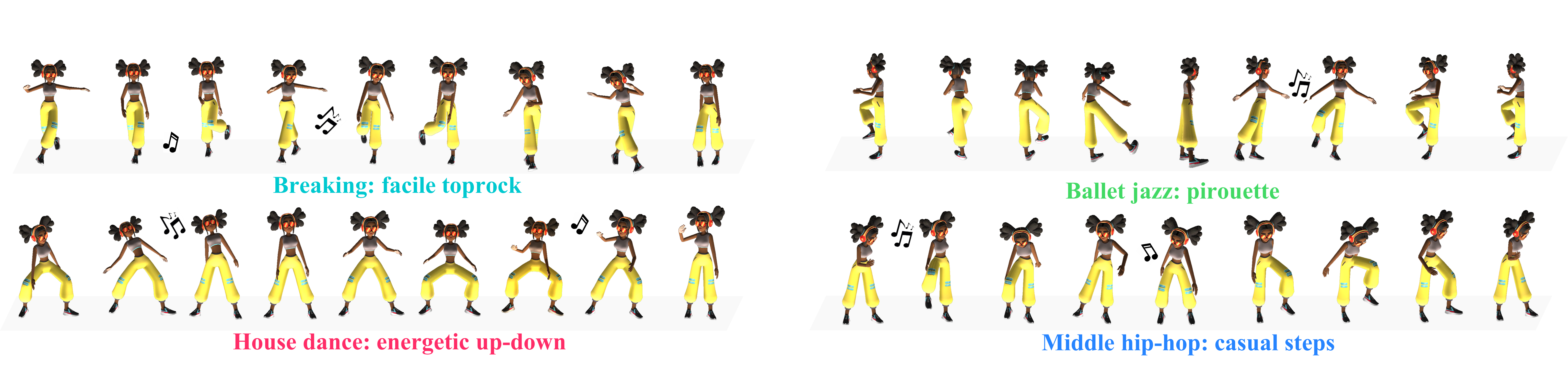}
        
    \vspace{-14pt}
    \captionof{figure}{\small{\textbf{Dance examples generated by our proposed method} on various types of music. The character is from Mixamo~\cite{mixamo}}} 
    \label{fig:teaser}
    }
\end{center}%
}]
\vspace{-3pt}

\maketitle

\newcommand\blfootnote[1]{%
\begingroup
\renewcommand\thefootnote{}\footnote{#1}%
\addtocounter{footnote}{-1}%
\endgroup
}

\blfootnote{\textsuperscript{\Letter}~Corresponding author}


\vspace{-2pt}
\begin{abstract}
\vspace{-2pt}

Driving 3D characters to dance following a piece of music is highly challenging due to the \textbf{spatial} constraints applied to poses by choreography norms. In addition, the generated dance sequence also needs to maintain \textbf{temporal} coherency with different music genres.
To tackle these challenges, we propose a novel music-to-dance framework, \textbf{Bailando}, with two powerful components: \textbf{1)} a choreographic memory that learns to summarize meaningful dancing units from 3D pose sequence to a quantized codebook, \textbf{2)} an actor-critic Generative Pre-trained Transformer (GPT) that composes these units to a fluent dance coherent to the music.
With the learned choreographic memory, dance generation is
realized on the quantized units that meet high choreography standards, such that the generated dancing sequences are confined within the spatial constraints.
To achieve synchronized alignment between diverse motion tempos and music beats, we introduce an actor-critic-based reinforcement learning scheme to the GPT  with a newly-designed beat-align reward function. 
Extensive experiments on the standard benchmark demonstrate that our proposed framework achieves state-of-the-art performance both qualitatively and quantitatively. 
Notably, the learned choreographic memory is shown to discover human-interpretable dancing-style poses in an unsupervised manner.
Code and video demo are available at \url{https://github.com/lisiyao21/Bailando/}.

\end{abstract}

\vspace{-10pt}









\begin{figure*}[t]
    \scriptsize
    \setlength{\tabcolsep}{1.5pt}
    \centering
    \small{
    \vspace{-6pt}
    \includegraphics[width=0.9\linewidth]{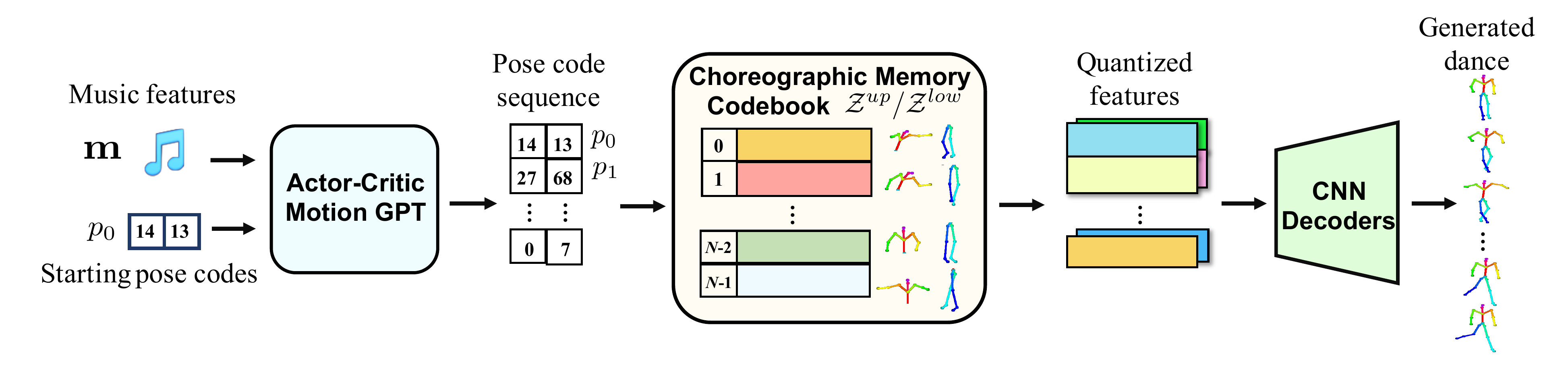}\\
    \vspace{-12pt}
    \caption{\small{\textbf{Dance generation pipeline of \textit{Bailando}.}  Given a piece of music, an actor-critic motion GPT autoregressively predicts the future upper-lower pose code pairs according to the music features and starting pose codes. The pose code sequence is then embedded to quantized features via a learned choreographic memory and finally decoded into a dance sequence by a CNN-based decoder.   }}
    \label{fig:pipeline}}
    \vspace{-13pt}
\end{figure*}

\vspace{-6pt}
\section{Introduction}
\vspace{-4pt}
\label{sec:intro}

Music-conditioned 3D dance generation
is an important task for its huge potential to facilitate a variety of real-world applications, \eg, assisting human artists choreograph and driving virtual characters performance. 
However, to produce satisfactory dancing sequence on given music is still very difficult due to two main challenges:
%
\textbf{1) Spatial constraint:} Not all the physically feasible 3D human poses are applicable for dance. The \textit{subspace} of dancing-style poses has stricter positional standards on body, and is selective to be visually expressive and emotionally infectious based on the choreography norms. 
%
\textbf{2) Temporal coherency with music:} The generated dancing sequence should be consistent with the music rhythm on various  genres of beats, while keeping the whole movements fluent.

Most existing dance generation studies intend to solve the two challenges both in a single ingeniously designed network that directly maps  music to 3D joint sequence in high-dimensional continuous space~\cite{Alemi2017GrooveNetR,Kao2020TemporallyGM,Tang2018DanceWM,Ginosar2019LearningIS,Ahn2020GenerativeAN,Li2021LearnTD}.
%
%
However, such methods are usually unstable in practice and are prone to regress to nonstandard poses beyond the dancing \emph{subspace}, \eg, freezing or meaningless swaying. Because there is no explicit constraints on target domain to restrict the synthesized dance to be spatially qualified.
To deal with the spatial constraint, some works 
collect real dancing clips as \emph{dance unit} and choreograph by splicing these units \cite{Ye2020ChoreoNetTM,choreomaster2021}.
While these methods guarantee the spatial quality of generated dance by directly manipulating on real data, the collection of dance units costs tremendous manual efforts, and they are not compatible with different rhythms. In addition, the units cannot be reused for different kinds of music beats due to their fixed length and speed.

In view of the shortcomings of existing methods, we propose a novel dance generation framework, \textbf{\textit{Bailando}}, that possesses two main components aiming at the spatial and temporal challenges, respectively.
First, to address the spatial challenge, a finite dictionary of quantized dancing units, namely \emph{choreographic memory}, is made by summarizing fundamental and reusable constituents from  movements in the dancing-style subspace.
Instead of manually indicating the dance units, we leverage the recent advances of VQ-VAE~\cite{Oord2017NeuralDR} to encode and quantize 3D joint sequence to a codebook in an unsupervised manner, where each learned code is shown to represent a unique dancing pose.
To further enlarge the range that choreography memory can represent, we divide 3D poses into compositional  upper and lower half bodies and learn VQ-VAEs for the half bodies separately,
such that any piece of dance can be represented into a sequence of paired pose codes.

Second, to generate temporally harmonic dance sequence, a GPT-like~\cite{radford2019language} network, named motion GPT, is introduced to translate music and source pose codes to targeted future pose codes.
%
Since the 3D poses are divided into compositional half bodies in the choreographic memory, we enhance our motion GPT with  proposed cross-conditional causal attention layer to keep the coherence of the generated body.
Moreover, to achieve accurate temporal synchronization between diverse motion tempos and music beats, we apply an on-policy reinforcement learning scheme to further improve the motion GPT via actor-critic~\cite{konda2000actor} finetuning with a newly-designed beat-align reward function.
%

The inference procedure of {\textit{Bailando}} is shown in Figure~\ref{fig:pipeline}.
Given a piece of music and a starting pose code pair, the actor-critic GPT autoregressively predicts the future pose code sequence,
which are then embedded to corresponding quantized features in choreographic memory, and are finally decoded and composed to 3D dance sequence by the dedicated CNN-based decoders of learned pose VQ-VAE.
The contributions of our work can be summarized in three folds:
\textbf{1)} A choreographic memory is created to encode and quantize dancing-style 3D poses, which is achieved by VQ-VAE in an unsupervised manner.
\textbf{2)} To align diverse motion tempos with different genres of music beats, an actor-critic GPT incorporated with the choreographic memory and cross-conditional causal attention is introduced. 
\textbf{3)} Extensive experiments show that our proposed \textit{Bailando} significantly outperforms the existing state of the art on both automatic metrics and visualization judgements. 
Code and models will be released upon acceptance.

\begin{figure*}[h]
    \scriptsize
    \setlength{\tabcolsep}{1.5pt}
    \centering
    \includegraphics[width=0.9\linewidth]{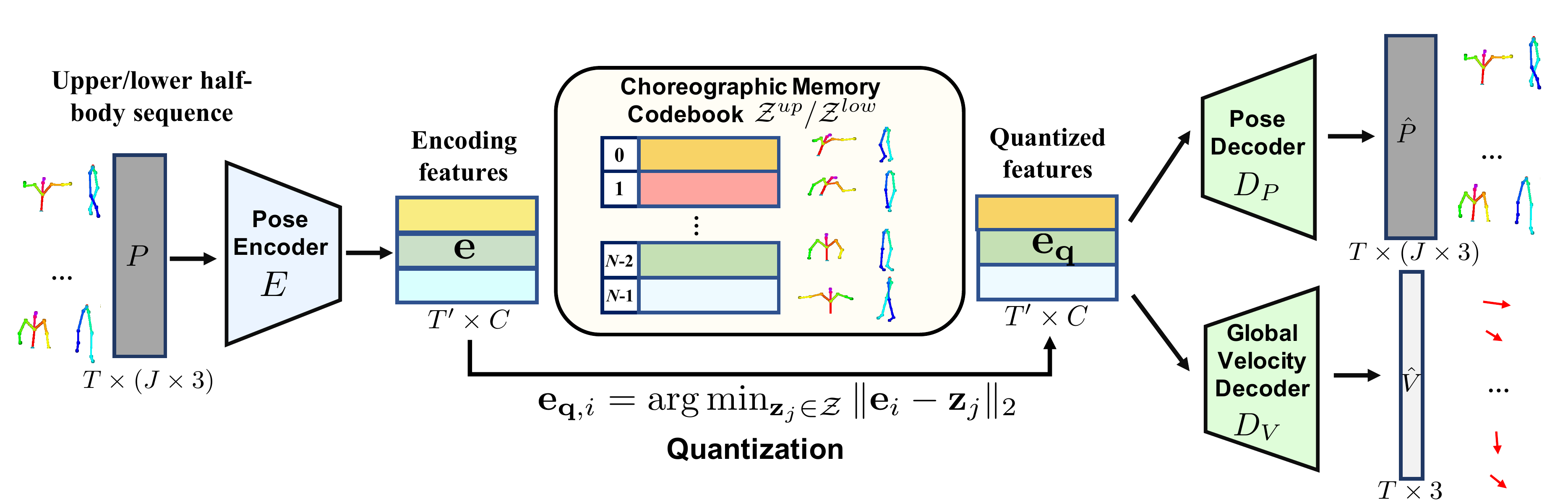}\\
    \vspace{-10pt}
    \caption{\small{\textbf{Structure of 3D Pose VQ-VAE.} 
    The proposed 3D pose VQ-VAE is learned to encode and summarize meaningful dancing units to choreographic memory, and to reconstruct the target pose sequence from quantized features.
    The parameters of encoder and decoders and the codebook are jointly learned during training.
    }}
    \label{fig:VQVAE}
    \vspace{-14pt}
\end{figure*}

\vspace{-4pt}
\section{Related Work}
\label{sec:related}
\vspace{-4pt}

\noindent{\textbf{Motion Synthesis and Music to Dance.}}
Producing realistic human motions has been long studied.
A typical class of approaches is \emph{graph-based} methods. They are developed on the idea of ``cropping and pasting'', which cut motion clips from existing data as  individual nodes and splice these nodes to synthesize new motions according to proper rules \cite{lamouret1996motion,arikan2002interactive,kovar2008motion,lee2002interactive}.
For music to dance, further constraints on the music rhythms, including source-target music similarity \cite{lee2013music}, beat-wise motion connectivity \cite{fukayama2015music},  and deep rhythm signatures \cite{choreomaster2021}, are introduced into the linking rules of the graph-based methods to align the motion with music beats. 
However, since the tempos, length, and speed of the cropped dance units are fixed, the graph-based methods would encounter temporal conflicts on diverse rhythms.
For example, the dance units cropped in music of  4/4 time signature cannot synthesize movement for 3/4, while the motion tempos of 60 beats per minute (BPM) is not adaptable for 80 BPM.
As a result, this kind of works can perform well in restricted rhythm ranges but is not compatible with various genres of music beats in wild scenarios.
%
%
In recent years, with the emergence of deep learning, many works design a dedicated network structure, including CNNs \cite{holden2016deep}, RNNs \cite{Tang2018DanceWM,Alemi2017GrooveNetR,yalta2019weakly,Huang2021DanceRL}, GCNs \cite{yan2019convolutional,ren2020self,Ferreira2021LearningTD}, GANs \cite{Lee2019DancingTM,sun2020deepdance} and Transformers \cite{Li2020LearningTG,Li2021LearnTD,Li2021DanceNet3DMB}, to map the given music to a joint sequence of the continuous human pose space directly.
Due to lacking explicit restrictions to keep the generated pose within the spatial constraint, such methods would regress to nonstandard poses that are beyond the dancing \emph{subspace} during inference, resulting in instability in real uses.
%
%
%
Besides various kinds of methods, different 3D dancing sequence data are made from mocap and reconstruction \cite{Tang2018DanceWM,Alemi2017GrooveNetR,Zhuang2020Music2DanceMD}. 
Recently, a large-scale 3D dancing dataset AIST++ \cite{Li2021LearnTD} is built from multi-camera videos along with the music in different styles and speeds, facilitating both training and testing of this task.

\noindent{\textbf{Two-Stage Generation.}}
The two-stage approaches,  which first encode data and afterwards learn a probabilistic model to generate the encoding, have been applied in multiple generative areas \cite{dhariwal2020jukebox,yan2021videogpt,esser2021taming}.
For example,  Dhariwal \etal~\cite{dhariwal2020jukebox} extracts audio features and generate songs according to the lyrics, while most recently Esser \etal~\cite{esser2021taming} encode perceptually rich image constituents to quantized patches and tames the Transformer to generate contextually plausible images in large resolutions.
In our work, we encode and quantize meaningful dancing constituents into a choreographic memory and generate visually satisfactory dance by jointly translating the music and existing movements to targeted future poses.


\vspace{-6pt}
\section{Our Approach}
\label{sec:method}
\vspace{-3pt}

The overview of our dance generation framework, \textit{{Bailando}}, is shown in Figure \ref{fig:pipeline}. 
Unlike other learning based methods, we do not learn a direct mapping from audio features to the continuous domain of 3D joint sequence.
%
Instead, we first encode and quantize the spatially standard dance movements into a finite codebook $\mathcal Z = \{{\bf z}_i\}_{i=0}^{N-1}$ as choreographic memory in Section~\ref{subsec:VQ-VAE}, where $N$ is the codebook length and every code ${\bf z}_i$ is shown to represent a dancing-like pose with contextual semantic
information.
Specifically,
we learn VQ-VAEs on the upper and lower half bodies separately, and represent the dance movement into a sequence of compositional upper-and-lower pose code pairs $\bm p=[p^{u}, p^{l}]$.
%
%
Then, we introduce a motion GPT to translate the music feature and source pose codes to the future pose codes in Section~\ref{subsec:ccgpt}.
Furthermore, to achieve synchronized alignment between generated motion tempos and music beats, we propose actor critic learning on the motion GPT with our newly designed beat-align rewards in Section~\ref{subsec:ac}.
%
The generated pose code sequences are finally decoded composed to fluent 3D dance by VQ-VAE decoders.

%

\vspace{-2pt}
\subsection{3D Pose VQ-VAE with Choreographic Memory}
\label{subsec:VQ-VAE}
\vspace{-2pt}

\begin{figure*}[h]
    \scriptsize
    \setlength{\tabcolsep}{1.5pt}
    \centering
    \small{
    \includegraphics[width=0.95\linewidth]{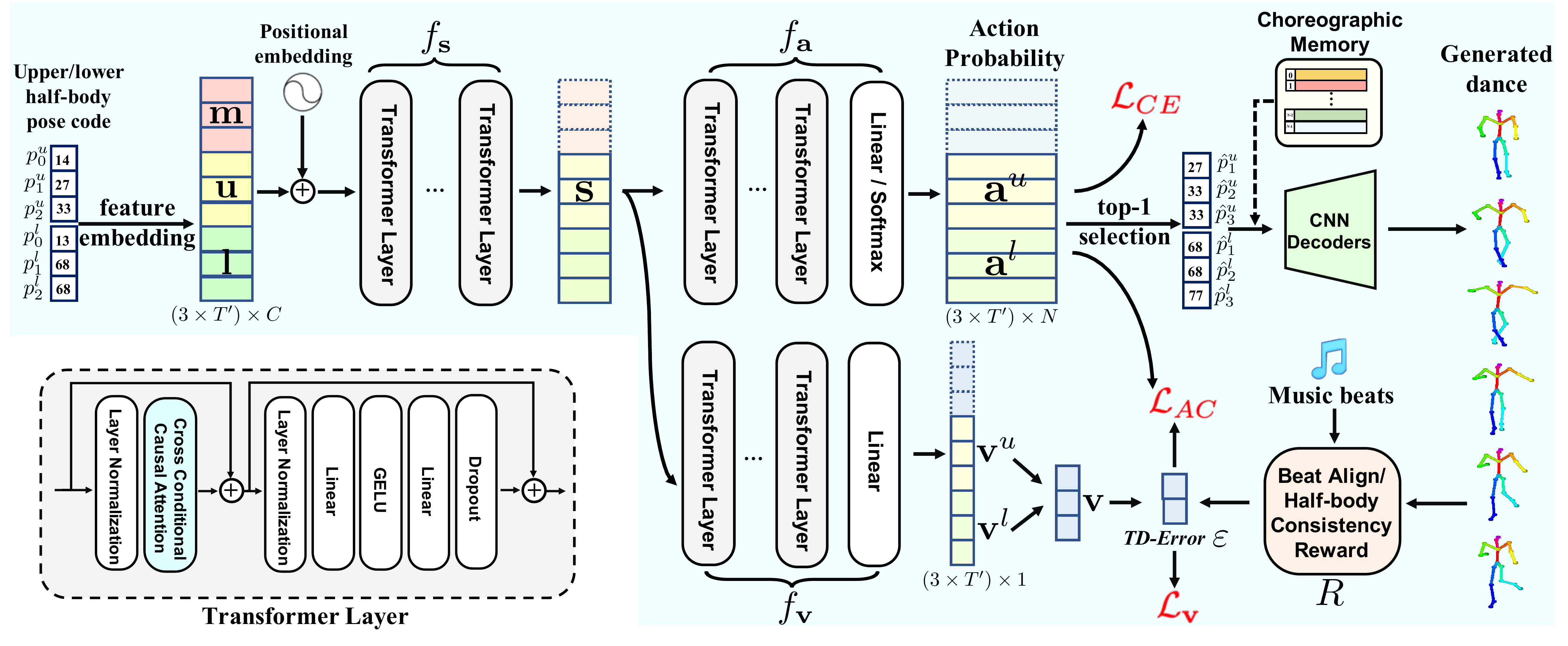} 
    \vspace{-8pt}
    \caption{\small{\textbf{Actor-Critic GPT.}  The GPT is learned to sequentially translate the source pose codes $[p^u_t, p^l_t]$ of upper-and-lower half bodies along with music features $\bf m$ to the targeted future pose codes $[\hat p^u_{t+1}, \hat p^l_{t+1}]$. The parameters of the networks are learned via cross-entropy loss $\mathcal L_{CE}$ with ground truth and actor-critic loss $\mathcal L_{AC}$.}}
    \label{fig:transformer_ac}}
    \vspace{-14pt}
\end{figure*}

Dance positions, \ie, the meaningful poses in dancing movements, 
are the basic constituents of a piece of dance and the process of choreography can be regarded as the combinations and connections of dance positions.
Although dances can vary greatly in style or speed, they share common dance positions.
Instead of indicating fixed units of dance with plenty of manual efforts, our goal is to
 summarize such dance positions into a rich and reusable codebook in an unsupervised manner, 
such that any piece of dance $P\in \mathbb R^{T\times(J\times3)} $, where $T$ is the time length and $J$ is joint amount, can be represented by a sequence of codebook elements ${\bf e}_{\bf q}\in \mathbb R^{T'\times C}$, where $T' = T/d$, $d$ is the temporal down-sampling rate, and $C$ is the channel dimension of features. 

To collect distinctive pose codes as well as to reconstruct them back to represented dancing sequence efficiently, we design a 3D pose VQ-VAE as shown in Figure~\ref{fig:VQVAE}.
In this scheme, we first adopt a 1D temporal convolution network $E$ to encode the 3D joint sequence $P$ to context-aware features ${\bf{e}} \in \mathbb R^{T'\times C}$.
%
Then, we quantize $\bf e$ by substituting each temporal feature ${\bf e}_i$ to its closest codebook element ${\bf z}_j$ as 
\begin{equation}
    {\bf e}_{{\bf q},i} = \arg\min_{{\bf z}_j \in \mathcal Z}{\|{\bf e}_i - {\bf z}_j\|}.
\end{equation}
%
%
Finally, we decode the quantized features  ${\bf e}_{\bf q}$ via a CNN $D_P$ and reconstruct the dance movement $\hat P$.

\noindent{\textbf{Compositional Human Pose Representation}}. In order to represent a larger range of motions by training on limited dance data, we train independent 3D pose VQ-VAEs and learn two separate codebooks $\mathcal Z^{u}$ and $\mathcal Z^{l}$ for the upper and lower half bodies, respectively, such that we can combine different upper-lower code pairs to enlarge the range of dance positions that the learned codebooks can cover.
Meanwhile, to avoid encoding confusion caused by global shift of joints (\eg, the same motion may be encoded to different features when it is at different locations), we normalize the absolute locations of input $P$, \ie, setting the root joints (hips) to be 0. To realize the overall movement, we add a separate decoder branch $D_V$, which predicts the global movement velocity $\hat V\in \mathbb R^{T\times 3}$  according to pose codes of the lower half body,  where $\hat V_t$ represents the shift of root joint between the $(t+1)$-th and the $t$-th frames.

\noindent{\bf Learning Stable 3D Pose VQ-VAEs.}
The pose encoder $E$ and decoder $D_P$ are simultaneously learned with the codebook via the following loss function:
\begin{equation}
\label{eq:quantize}
    \mathcal L_{VQ} = \mathcal L_{rec}(\hat P,P) + \|\text{sg}[\bf e] - \bf e_{\bf q}\| + \beta \|{\bf e} - \text{sg}[{\bf e_{\bf q}}]\|.
\end{equation}
The global velocity decoder branch is learned afterwards by fixing the parameters of other parts of VQ-VAE via loss function $\mathcal L_{rec}(\hat V,V)$, where $V$ is the ground truth global velocity.
$\mathcal L_{rec}$ is the reconstruction loss that constrains the predicted 3D joint sequence to ground truth. In this loss, we regress not only the original 3D points of joints,  but also the velocities and accelerations of movements:
\begin{equation}
    \mathcal L_{rec}(\hat P, P) = \|\hat P -P\|_1 + \alpha_1\|\hat P' - P'\|_1 + \alpha_2\|\hat P'' - P''\|_1, 
\end{equation}
where $P'$ and $P''$ represent the 1$^{st}$-order (velocity) and 2$^{nd}$-order (acceleration)  partial derivatives of 3D joint sequence $P$ on time, while $\alpha_1$ and $\alpha_2$ are trade-off weights.
Experimental results show the ``velocity-and-acceleration'' loss items play essential roles to prevent jitters in generated dance. 
(See Section \ref{sec:ablation}.)

The second part of $\mathcal L_{VQ}$ is the ``codebook loss'' to learn codebook entries,
where $\text{sg}[\cdot]$ denotes ``stop gradient'' \cite{chen2021exploring},
%
while the third part is the ``commitment loss'' with trade off $\beta$ \cite{esser2021taming,dhariwal2020jukebox}.
Since the quantization operation of Equation~\ref{eq:quantize} is not differentable, to train the whole networks end to end, the back-propagation of this operation is achieved by simply passing the gradient of $\bf e_{\bf q}$ to $\bf e$.

The learned choreographic memory codes are interpretable.
After the training process of pose VQ-VAEs, 
each quantized feature in the codebook is decoded into a unique dance position. 
And any permutation and combination of  codes  can be decoded to a piece of fluent movement based on corresponding dance positions. (See Section~\ref{subsec:interpretable}.)

\subsection{Cross-Conditional Motion GPT}
\label{subsec:ccgpt}

%
\vspace{-2pt}

Now that we can represent any piece of dance by a sequence of quantized position codes, the dance generation task is then reframed as to select proper codes  from codebook $\mathcal Z$ for future actions according to given music and existing movements.
For any target time $t$, we estimate the probability of every ${\bf z}_i \in \mathcal Z$ and select the one with the largest possibility as the predicted pose code $\hat {\bm p}_t$.
Since we model the upper and lower half bodies separately, in order to keep the coherence of composed body and to avoid the  asynchronous situation (\eg, the direction of the upper half is opposite to that of the lower), the prediction of the future action should be cross-conditioned between existing upper and lower movements to make the most of mutual information:
%
\begin{equation}
\left\{ 
        \begin{array}{lll}
              \hat p^u_t &=& \arg\max_{k} {{\mathbb{P}}({\bf z}^{u}_k|{\bf m}_{1...t}, p^u_{0...{t-1}}, p^l_{0...t-1})} \\
              \hat p^l_t &=& \arg\max_{k} {{\mathbb{P}}({\bf z}^{l}_k|{\bf m}_{1...t}, p^u_{0...{t-1}}, p^l_{0...t-1})}
        \end{array}
        \right.
\end{equation}

%
We introduce the powerful GPT model \cite{radford2019language}
to estimate the action probabilities as shown in Figure~\ref{fig:transformer_ac}.
Given a dance position code sequence with length of $T'$, we first embed the upper and lower pose codes to learnable features ${\bf u}\in\mathbb R^{T'\times C}$ and ${\bf l}\in\mathbb R^{T'\times C}$, respectively, and concatenate them with music features $\bf m$ on the temporal dimension.
Then, we add a learned positional embedding to this concatenated $(3\times T')\times C$ tensor and feed it to 12 successive Transformer layers, the structure of which is shown in Figure~\ref{fig:transformer_ac}.
At last, we employ a linear transform  and softmax layer to map the output of Transformer layers to normalized action probability ${\bf a} \in \mathbb R^{(3\times T')\times N}$, where $N$ is the size of learned codebook and ${\bf a}_{t, i}$ reveals the probability of pose code ${\bf z}_i \in \mathcal Z$ predicted for time $t+1$. The action probabilities for upper and lower half bodies are indexed as ${\bf a}^u_{0:T'-1} = {\bf a}_{T':2T'-1}$ and ${\bf a}^l_{0:T'-1} = {\bf a}_{2T':3T'-1}$, respectively.
%

%


\begin{figure}[t]
    \scriptsize
    \setlength{\tabcolsep}{1.5pt}
    \centering
    \includegraphics[width=0.9\linewidth]{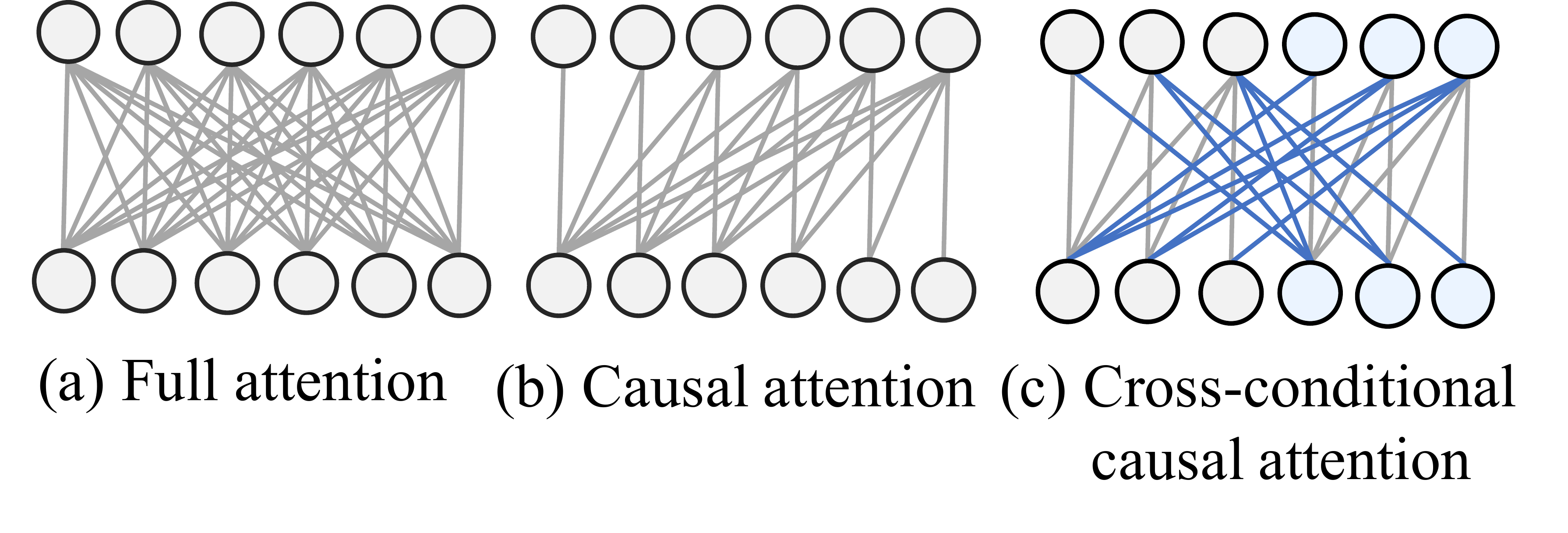}
    \vspace{-16pt}
    \small{
    \caption{\small{\textbf{Different types of attention layers.} The proposed cross-conditional causal attention realizes causal inferences intra (gray lines) and inter (blue lines) different kinds of components (gray and blue circles). Two kinds of components are shown here for concision, but three (music, upper, lower bodies) are in reality.} }
    
    \label{fig:attention}}
    \vspace{-14pt}
\end{figure}

In Transformers~\cite{NIPS2017_3f5ee243}, the attention layer is the core component that determines the computational dependency among sequential elements of data, and is implemented as
\begin{equation}
    \text{Attention}({\bf Q, K, V, M}) = \text{softmax}\left(\frac{{\bf QK}^T + {\bf M}}{\sqrt{C}}\right) \bf V,
\end{equation}
where $\bf Q, K, V$ denote the query, key and value from input, and $\bf M$ is the mask, which determins the type of attention layers. 
The most two common types of attentions are ``full attention''~\cite{NIPS2017_3f5ee243} and ``causal attention''~\cite{NIPS2017_3f5ee243}, where the former realizes the intercommunication of input data at all times while the latter only allows the current and previous data to compute the state for the time of interest.
As our goal is to infer the future dance position codes, we adopt the causal attention.
However, since the generation of upper and lower half bodies are dependent on each other, we cannot realize the inference by just reordering the sequence of input to fit the causality as previous works \cite{esser2021taming, dhariwal2020jukebox}.
Therefore, we propose an attention layer, namely cross-conditional attention, to comply the causality cross conditioned among features of the music, the upper half and the lower half bodies, where $\bf M$ is designed to be a $3\times3$ repeated block matrix with a lower triangular matrix of size $T'$ as its element.
As shown in Figure~\ref{fig:attention}, the proposed attention can exchange information of different components, and guarantee that the future information will not be transmitted back to the past.

\noindent{\bf Learning Motion GPT.}
The motion GPT is optimized via supervised training with cross-entropy loss on action probability $\bf a$:
\begin{equation}\label{eq:loss_ce}
    \mathcal L_{CE} = \frac{1}{T'}\sum_{t=0}^{T'-1} \sum_{h =u, l} {\text{CrossEntropy}\left({\bf a}^h_t, p^h_{t+1}\right)}.
\end{equation}
%

%
Given a sequence of pose codes $\bm p_{0:T'-1}$  and relevant music features ${\bf m}_{1:T'}$ as input, the learned GPT outputs the sequence of actions ${\bf a}_{0:T'-1}$ all at once to predict $\bm p_{1:T'}$.
This parallel characteristic makes Transformer an ideal  model for reinforcement learning \cite{janner2021reinforcement,chen2021decision}.
In the following subsection, we adopt the learned motion GPT as a pretrained policy maker and propose a novel actor-critic based finetuning scheme to further improve its performance as complementary to the supervised training above.
%
\subsection{Actor-Critic Learning}

While the supervised learning scheme for the motion GPT is straightforward and easy to train, it is intractable to further involve a more flexible constraint of generated dance (\eg, a regularization item that strengthens the consistency of dance beats) to Equation~(\ref{eq:loss_ce}), since the supervision target is the code number, which is not differentiable to compute the quantitative constraints on the final dance sequence.

To address this issue and to achieve more accurate synchronized alignment between diverse motion tempos and music beats, we 
apply  actor-critic learning to the motion GPT with a newly-designed reward function.
In particular, we regard the first 6 Transformer layers of motion GPT as ``state network'' $f_{\bf s}$, and the outputs of $f_{\bf s}$ are states ${\bf s}$ for time $0$ to $T'-1$, while the latter 6 Transformer layers along with the linear-softmax layer are regarded as ``policy making network'' $f_{\bf a}$, where the actions are computed according to state as ${\bf a} = f_{\bf s}({\bf s})$.
%
Besides, we add a separate three-layer Transformer branch as ``critic value network'' $f_{\bf v}$ to estimate the critic values ${\bf v}_{0:T'-1} \in \mathbb R^{T'\times 1}$ as 
\begin{equation}
    {\bf v} = {\bf v}^u + {\bf v}^l = f_{\bf v}({\bf s})_{T':2T'-1} + f_{\bf v}({\bf s})_{2T':3T'-1}. 
\end{equation}

With well defined reward function $R(t)={R({\bf a}_t, {\bf s}_t)}$, the objective of reinforcement learning is to maximize the expected accumulated rewards:
\begin{equation}
    J = \mathbb E_{\bf \tau}\left[\sum_{t=0}^{T'-1}{R(t)}\right],
\end{equation}
where ${\bf \tau} = \{{\bf a}_t\}_{t=0}^{T'-1}$ is the trajectories of actions predicted by the policy making network.
This objective is then converted to optimize the parameters of policy making network using the following loss function:
\begin{equation}\label{eq:loss_ac}
\begin{aligned}
& \mathcal L_{AC} = \\ 
&\,\,\,\frac{1}{T'-1} \sum_{t=0}^{T'-2} \left(  \sum_{h =u, l}{\text{CrossEntropy}\left({\bf a}^h_t, \hat p^h_{t+1}\right)} \right)\cdot \text{sg}[{\bm \varepsilon_t}],
\end{aligned}
\end{equation}
where $\hat p_{t+1}^h = \arg\max_i{{\bf a}^h_{t, i}}$ is the pose code number predicted by the policy making network. $\bm \varepsilon \in \mathbb R^{(T'-1)\times 1}$ denotes the so-called TD-error calculated as
\begin{equation}
\bm \varepsilon_{0:T'-2} = {\bf r}_{0:T'-2} + \text{sg}[{\bf v}_{1:T'-1}] -  {\bf v}_{0:T'-2},
\end{equation}
where ${\bf r}_t = R(t)$.
The detailed derivation of Equation (\ref{eq:loss_ac}) can be found in the supplementary file.
Meanwhile, the critic value network is optimized by bootstrap training on difference between ${\bf v}_{0:T'-2}$ and $R({\bf a}_t, {\bf s}_t) + {\bf v}_{1:T'-1}$:
\begin{equation}
\mathcal L_{\bf v} = \frac{1}{T'-1}\|\bm \varepsilon\|^2_2.
\end{equation}

The computation of actor-critic loss $\mathcal L_{AC}$ depends on real-time actions predicted by the motion GPT and the optimization direction is determined on the value of TD-error.
When $\varepsilon_t$ is positive, the optimization on $\mathcal L_{AC}$ will intensify the convergence to predicted code $\hat p_{t+1}$, while in the opposite situations, the   probability estimated for $\hat p_{t+1}$ will be reduced.

\begin{figure}[t]
    \scriptsize
    \setlength{\tabcolsep}{1.5pt}
    \centering
    \small{
    \includegraphics[width=0.9\linewidth]{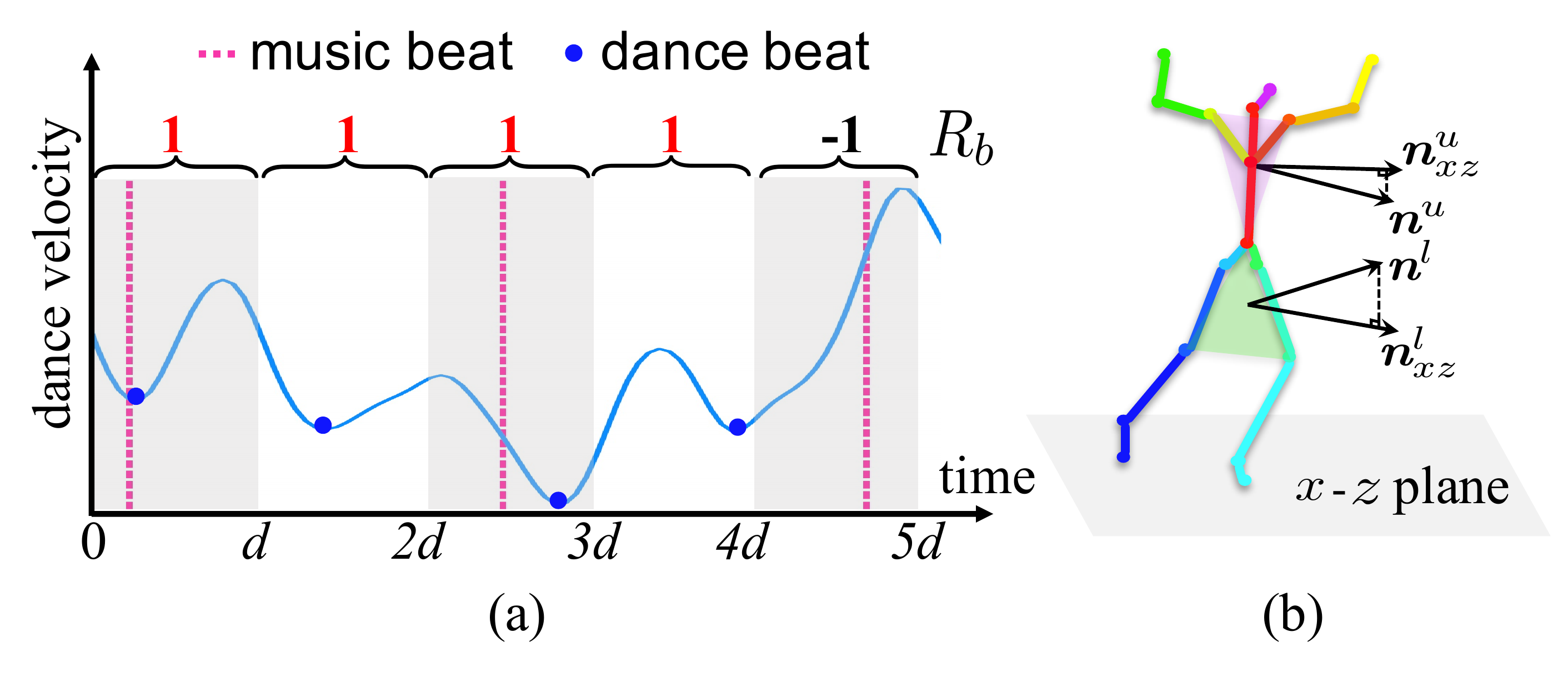}\\
    \vspace{-10pt}
    \caption{\small{\textbf{Designed rewards.}  (a) Beat-align reward penalizes the absence of dance beat for the interval that has music beat. (b) Half-body consistency reward is computed on the angle between normal directions of half bodies to prevent asynchronizations.}}
    
    \label{fig:rewards}}
    \vspace{-14pt}
\end{figure}

The value of TD-error and the learning effectiveness are strongly influenced by the reward function $R$.
In this work, we design a motion-music beat-align reward to generate dance more accurate to the rhythm of music.
As shown in Figure \ref{fig:rewards} (a), the beat-align reward is defined as
\begin{equation}
    R_b(t) = \left\{ 
        \begin{array}{lll}
              -1,&{\textit{\text{$\exists$ music beat $\land \nexists$ dance beats $\in \hat P_{td:(t+1)d}$}}} \\
              1,&\textit{\text{otherwise},}
        \end{array}
    \right.
\end{equation}
where $\hat P_{0:T-1} =  D(\hat{\bm p}_{0:T'-1})$ is the dance motion sequence decoded from predicted dance position codes.
Meanwhile, to avoid the compositional asynchronization between upper and lower half bodies during actor-critic learning, we introduce a compositional consistency reward to impose penalties for situations where the upper and lower body are in the opposite direction:
\begin{equation}
    R_c(t) = \inf\left\{\hat R_c(t)\right\}, t\in\left[dt, d(t+1)\right),
\end{equation}
where
\begin{equation}
        \hat R_c(t) = \left\{ 
        \begin{array}{lll}
              \left<\bm n^u_{xz}(t),  \bm n^l_{xz}(t)\right>,  & \left<\bm n^u_{xz}(t),  \bm n^l_{xz}(t)\right> < 0\\
              1,&\textit{\text{otherwise}.}
        \end{array}
    \right.
\end{equation}
Here, $\bm n^u_{xz}(t),  \bm n^l_{xz}(t)$ are the normal directions of upper and lower bodies of $\hat P_t$ projected to the $x$-$z$ plane, which is illustrated in Figure \ref{fig:rewards} (b).
The final reward is then a weighted combination of $R_b$ and $R_c$ as $R = \gamma_bR_b + \gamma_cR_c$.

In the finetuning process, we fix the parameters of state network $f_{\bf s}$, and  alternately train the policy making network $f_{\bf a}$ and the critic value network $f_{\bf v}$ using the losses introduced above with a small learning rate.
After such finetuning, the proposed framework will be further enhanced.

\label{subsec:ac}

\section{Experiments}
\vspace{-4pt}
\label{sec:experiment}

\begin{table*}
    
\caption{\textbf{Quantitative results on AIST++ test set.} The best and runner-up values are bold and underlined, respectively.
Among compared methods, ``Li \etal'', DanceNet and FACT are multiplexing the same results of AIST++ benchmark~\cite{Li2021LearnTD},  while DanceRevolution~\cite{Huang2021DanceRL} is reproduced using officially released code with the optimal settings.
$\dag$ FID$_k$ and DIV$_k$ are fetched from \cite{Li2021LearnTD} while FID$_g$ and DIV$_g$ are recomputed using the officially updated evaluation code.
%
*The generated dances of ``Li \etal'' are highly jittery making its velocity variation extremely high, which is also reported in \cite{Li2021LearnTD}.
}
\vspace{-8pt}
  
  \centering
  \small
{
  \begin{tabular}{l  c c c c c c   }
    \toprule
    &  \multicolumn{2}{c}{ Motion Quality} & \multicolumn{2}{c}{ Motion Diversity } &  & \multicolumn{1}{c}{ User Study}             \\

     \cmidrule(r){2-3} \cmidrule(r){4-5}  \cmidrule(r){7-7} 
   Method & FID$_k\downarrow$ & FID$_g^{\dag}\downarrow$ &  Div$_k\uparrow$ &   Div$_g^{\dag}\uparrow$ &  Beat Align Score $\uparrow$  & Our Method Wins \\
      \midrule

Ground Truth & 17.10 & 10.60 & 8.19 & 7.45 & 0.2374 & 40.0\%$\pm$25.2\%\\
\midrule
Li \etal~\cite{Li2020LearningTG} & 86.43 & 43.46 & 6.85$^*$ & 3.32 & 0.1607 & 100.0\%$\pm$0.0\%$\,\,\,\,\,\,$\\
DanceNet~\cite{Zhuang2020Music2DanceMD} & 69.18 & 25.49  & 2.86 &  2.85 & 0.1430 &  92.7\%$\pm$12.1\%\\
DanceRevolution~\cite{Huang2021DanceRL} & 73.42 & 25.92 & 3.52 & 4.87 & 0.1950 & 84.5\%$\pm$10.8\% \\
FACT~\cite{Li2021LearnTD} & \underline{35.35} & \underline{22.11} & \underline{5.94} & \underline{6.18} & \underline{0.2209} & 98.2\%$\pm$3.9\%$\,\,\,$\\


\rowcolor{mygray}
{{\textit{Bailando}}} (Ours) & \bf 28.16 & \bf 9.62 & \bf 7.83 & \bf 6.34 & \bf{0.2332} & -- \\

    \bottomrule
  \end{tabular}
  }
  \label{table:quantitative_eval}
   \vspace{-15pt}
\end{table*}


\noindent{\textbf{Dataset}.}
We perform the training and evaluation on the AIST++ dataset proposed in \cite{Li2021LearnTD}, which to our best knowledge is the largest public available dataset for paired music and motions. 
This dataset contains 992 pieces of high-quality $60$-FPS 3D pose sequence in SMPL format \cite{SMPL:2015}, where 952 are kept for training and 40 are used for evaluation.
%
%

\noindent{\textbf{Implementation Details}.}
In this work, the choreographic memory codebook size $N$ for both upper and lower bodies is set to $512$, while the channel dimension $C$ of encoded features is $512$ and the temporal downsampling rate $d$ of encoders is $8$.
The structures of the convolutional encoder and decoders are provided in the supplementary file.
While training the VQ-VAEs,  dance data are cropped to length of $T=240$ ($4$ seconds)  and sampled in batch size of $32$.
The commit loss trade-off $\beta$ in $\mathcal L_{VQ}$ is $0.1$, while  $\alpha_1$ and $\alpha_2$ in $\mathcal L_{rec}$ are both set to be $1$.
We adopt Adam optimizer \cite{kingma2014adam} with $\beta_1=0.9$ and $\beta_2=0.99$ to train both pose VQ-VAEs for $400$ epochs with learning rate $3\times10^{-5}$.
As to the motion GPT, we comply a structure mirroring \cite{mingpt}, where the channel dimension is $768$, and the attention layer is implemented in $12$ heads with  dropoout probability $0.1$.
The music features are extracted by the public audio processing toolbox Librosa \cite{jin2017towards}, including \emph{mel frequency cepstral
coefficients (MFCC)}, \emph{MFCC delta}, \emph{constant-Q chromagram}, \emph{tempogram} and \emph{onset strength}, which are $438$-dim in total, and are mapped to the same dimension of GPT via a learned linear transform.
The block size $T'$ of GPT is set to be $29$. 
While training, the dance sequences are first encoded to pose codes $\bm p$ and sampled to length of $30$, where $\bm p_{0:28}$ are used as input and $\bm p_{1:29}$ are supervision labels.
The motion GPT is optimized using Adam optimizer with $\beta_1=0.5$ and $\beta_2=0.99$ for $400$ epochs, where the learning rate is initialized as $3\times10^{-4}$ and decayed after $200$ epochs with factor $0.1$.
In the actor-critic finetuning process, we adopt a small learning rate of $1\times 10^{-5}$ to learn $f_{\bf a}$ and $f_{\bf v}$ for $10$ epochs.
The reward trade-offs $\gamma_b$ and $\gamma_c$ are $5$ and $1$, respectively.
In our experiment, the pose VQ-VAEs and the motion GPT are trained sequentially, and the weights of VQ-VAEs are fixed during the learning process of GPT.
The whole framework is learned in three days on one Tesla V100 GPU.
During test, the motion GPT takes a pair of starting pose codes, which can be either manually indicated or randomly sampled, as input and autoregressively generates the motion sequence as long as the target music.

\noindent{\bf Evaluation Metrics.}
For quantitative evaluations, we measure the generated dance from three perspectives: the quality of generated dances, the diversity of motions and the alignment between the rhythms of music and generated movements.
In concrete, for the dance quality, we calculate the Fréchet Inception Distances (FID) \cite{heusel2017gans} between the generated dance and all motion sequences (including training and test data) of the AIST++ dataset on kinetic features \cite{onuma2008fmdistance} (denoted as `$k$') and geometric features \cite{muller2005efficient} (denoted as `$g$'), which are both extracted using the toolbox of \cite{gopinath2020fairmotion}.
As to the diversity, we compute the average feature distance of generated movements following \cite{Li2021LearnTD}.
Regarding to the alignment between music and generated motions, we calculate the average temporal distance between each music
beat and its closest dance beat as the Beat Align Score:
\begin{equation}
    \frac{1}{|B^m|}\sum_{t^m \in B^m}\exp{\left\{-\frac{\min_{ t^d \in B^d} {\|t^d - t^m\|^2}}{2\sigma^2}\right\}},
\end{equation}
where $B^d$ and $B^m$ record the time of beats in dance and music, respectively, while $\sigma$ is normalized parameter which is set to be $3$ in our experiment.

\vspace{-3pt}
\subsection{Comparison to Existing Methods}
\vspace{-3pt}

We compare our proposed model
to several state-of-the-art methods including Li \etal \cite{Li2020LearningTG}, DanceNet \cite{Zhuang2020Music2DanceMD}, DanceRevolution \cite{Huang2021DanceRL} and FACT \cite{Li2021LearnTD}.
For each method, we generate $40$ pieces of dances in AIST++ test set,
and sample the generated dance sequence with length of $20$ seconds to compute the evaluation metrics mentioned above.
We also calculate the quantitative scores for ground truth data in AIST++ test set and compare it to the generated dances.
%
%

%
The quantitative results are shown in Table~\ref{table:quantitative_eval}.
According to the comparison, our proposed model consistently performs favorably against all the other existing methods on all evaluations.
Specifically, our method improves $7.19$ ($20\%$) and $12.49$ ($56\%$) than the best compared baseline model FACT on FID$_k$ and FID$_g$, respectively, and even achieves a better FID$_g$ score than the ground truth ($9.62$ v.s. $10.60$).
%
%
If look closely to the metrics on these two kinds of features, the kinetic feature is defined on motion velocities and energies, which reflects the physical characteristics of dance, while the geometric feature is defined based on multiple man-made templates of movements, which reflects the quality of choreography.
The superiority of our method on both dance quality metrics reveals that {{\textit{Bailando}}} not only synthesizes more real-like motions than the compared baseline methods, but also achieves outstanding performance on organizing the movements to dance via the proposed  actor-critic GPT  scheme with learned choreographic memory.
Meanwhile, {{\textit{Bailando}}} can generate dance with high choreographic diversity instead of converging to few templates, and also achieves improvement on the correlation between music and motion.

\noindent{\bf User Study.}
To further understand the real visual performance of our method, we conduct a user study among the dance sequences generated by each compared method and the ground truth data in AIST++ test set.
The experiment is conducted with 11 participants separately.
For each participant, we randomly play 50 pairs of  comparison videos with a length of around 10 seconds, where each pair contains our result and one competitor's in the same music, and ask the participant to indicate ``\textit{which one is dancing better to the music}''.
The statistics are shown in Table~\ref{table:quantitative_eval}.
%
%
Notably, our method significantly surpasses the compared state-of-art methods with at least $84.5\%$ winning rate.
%
%
Even in comparison to the ground truth, $40\%$ of our generated dance is voted as the better in average.
According to the feedback from participants, our generated dance is more ``stable to the rythm'' with ``higher diversity'', while the reason why our method is still not as good as real dances is mainly due to ``lacking of long-term regularity  and subjective beauty''.
A detailed winning rate distribution on styles of dance can be referred to the supplementary file.

%

%

\vspace{-4pt}
\subsection{Ablation Studies}
\label{sec:ablation}
\vspace{-3pt}

We conduct ablation studies on the pose VQ-VAEs and the motion GPT, respectively.
The quantitative scores are shown in Table \ref{table:ablation}.
The visual comparisons of this study can be also referred to the supplementary video.
%

\noindent{\bf Pose VQ-VAE.}
We explore the effectiveness of the following components: (1) the up-lower half body separation, (2) the global velocity prediction branch, and (3) the velocity-and-acceleration loss used in $\mathcal L_{rec}$.
We train three variant models without each of the three components, respectively.
%
%
The motion quality measured for VQ-VAEs is on reconstructed results of ground truth of AIST++ test set.
%
As shown in Table~\ref{table:ablation}, the FID$_k$ and FID$_g$ values for variant ``w/o. upper/lower'' become worse by $12.98$ ($46\%$) and $3.22$ ($25\%$), respectively.
%
The VQ-VAE trained on whole body cannot reconstruct the dancing pose of test set effectively.
%
%
%
%
Therefore, the separate representations of upper-lower half bodies are necessary to enlarge the range of poses that the choreographic memory can cover.
%
As to the global velocity branch, the motion quality scores of ``w/o. global vel.'' sharply drops $42.72$ ($151\%$) and $5.89$ ($47\%$), respectively, which shows the isolated velocity prediction is critical for representing the dance movement.
For ``w/o. vel./acc. loss'' variant, the FID$_k$ is worsened by $2.68$.
Although the FID$_g$ value of ``w/o. vel./acc. loss'' is slightly improved by $0.76$, 
the model produces strong motion jitters if without adopting vel./acc loss for training in the supplementary video.

\begin{table}
    \centering
    \small{
\caption{\small{\textbf{Ablation study on AIST++ test set} {Experiments are conducted on pose VQ-VAE and  GPT, respectively.} }}
\vspace{-10pt}
  
  \centering{
  \begin{tabular}{l l  c c c}
    \toprule

 &  Method & FID$_k\downarrow$ & FID$_g\downarrow$ & BAS $\uparrow$\\
\midrule
\multirow{5}[0]*{\rotatebox{90}{Pose VQ-VAE}} & {Ground Truth} &  17.10 & 10.60 & -- \\
 
 & w/o. upper/lower  & 41.21  & 15.85  & --\\
 & w/o. global vel. & 70.95 & 18.52 & --\\  
& w/o. {vel./acc. loss}  & 30.91 & 11.87 & -- \\
&
\cellcolor{mygray}
 full pose VQ-VAE &  \cellcolor{mygray} 28.23 &  \cellcolor{mygray} 12.63 & \cellcolor{mygray} --\\

\midrule
\multirow{4}[0]*{\rotatebox{90}{GPT}} 
& w/o. quantization & 42.71 & 147.28 & --\\
& w/o. cross-cond. att.  & 37.41 & 15.52 & --\\
& w/o. actor critic & 28.75 & 11.82 & 0.2245\\
&
\cellcolor{mygray}
 full actor-critic GPT &  \cellcolor{mygray} 28.16 & \cellcolor{mygray} 9.62 & \cellcolor{mygray} {0.2332}\\
    \bottomrule
  \end{tabular}
  }
  \label{table:ablation}
  \vspace{-18pt}
  }
\end{table}

\noindent{\bf Motion GPT.}
%
For the proposed actor-critic GPT, first, we explore the effect of quantized choreography memory
by training a variant GPT 
directly 
regress to the encoding features of 3D joint sequence via an $L_2$ Loss.
As shown in Table \ref{table:ablation}, the FID$_g$ drops $135.41$ for variant ``w/o. quantization'' (compared to ``w/o. actor critic'', same below), while the generated dance sequences contain frequent jitters in vision, which shows the quantization of dancing positions is essential to our proposed framework.
Second, to test the effectiveness of the proposed cross-conditional causal attention, we substitute it to causal attention, and train two motion GPTs for upper and lower half bodies separately. 
The motion quality scores of ``w/o. cross-cond. att.'' drop $8.66$ ($30\%$) and $3.70$ ($31\%$), respectively. 
The main reason for the poor performance is that the generated dances of  contain frequent asynchronization of upper and lower half bodies, while the proposed cross-conditional attention layer can effectively prevent such situations via the interaction of information between the half bodies. 
%
%
At last, we compare the motion quality and music-motion consistency between the model with (denoted as ``full actor-critic GPT'') and without actor critic finetuning (denoted as ``w/o. actor critic'').
After the actor-critic learning, the beat-align score (BAS) of motion GPT increases from $0.2245$ to $0.2332$, proving the effectiveness of reinforcement learning scheme with proposed beat-align reward.
Meanwhile, by constraining the consistency with music, the actor-critic finetuning process can also enhance the motion quanlity on choreography and saliently improves the FID$_g$ score by $2.20$ ($19\%$).


\begin{figure}[t]
    \small{
    \scriptsize
    \setlength{\tabcolsep}{1.5pt}
    \centering
    \vspace{-5pt}
    \includegraphics[width=0.9\linewidth]{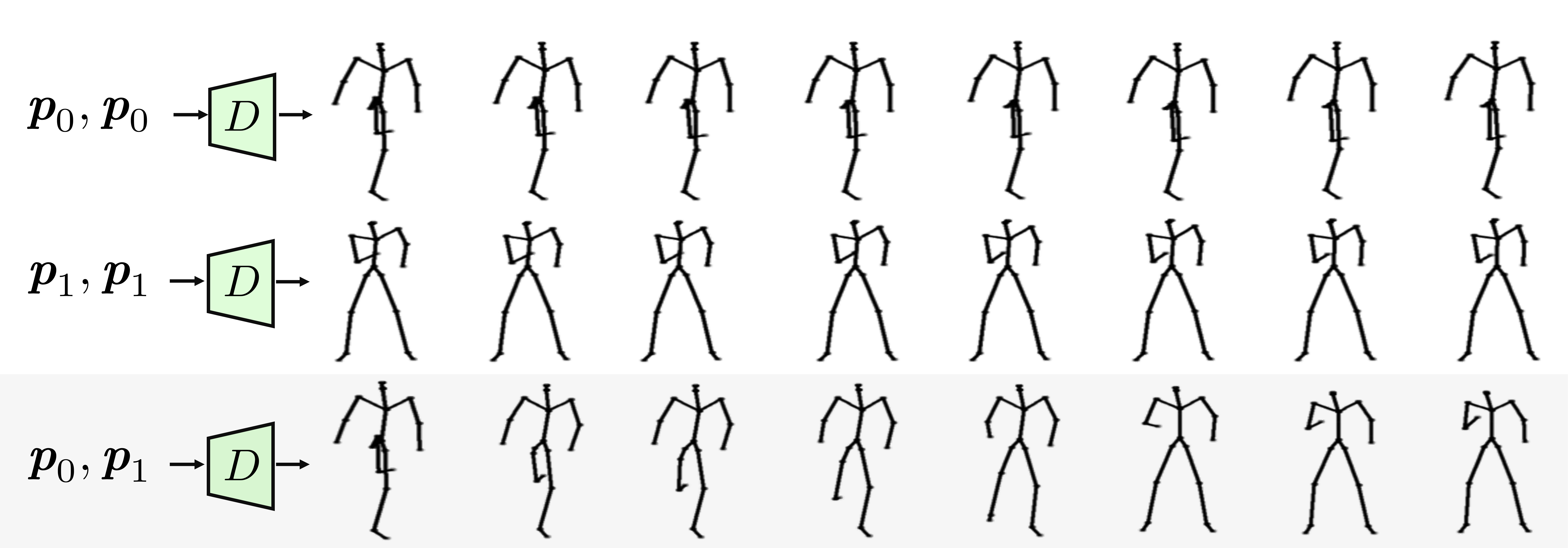}\\
    \vspace{-3pt}
    \small\caption{\small{\textbf{Interpretability of choreographic memory code.} The sequence of single code is decoded to static pose, while the sequence of two various codes is decoded to smooth transition between two poses, which means each code represents a dancing-style pose and the decoder links poses of different codes to movements.}}
    \label{fig:ablation}}
    \vspace{-14pt}
\end{figure}

\vspace{-5pt}
\subsection{Interpretability of Choreographic Memory}
\label{subsec:interpretable}
\vspace{-2pt}
In this work, we propose to  summarize meaningful dancing units into the codebook via pose VQ-VAE in an unsupervised manner.
To understand what kind of dance unit is learned in the choreographic memory, we visualize the latent codes and find each code represents a unique 3D dancing-style pose.
As revealed in Figure \ref{fig:ablation}, the first and the second rows are 3D poses decoded from $\bm p_0 = [4, 4]$ and $\bm p_1 = [5, 5]$, respectively, where the former one is doing right leg lifting and the latter is right bicep curl.
The decoded pose will keep static for repeating codes, and will make smooth transition between postures of different codes.
As shown in the third row of Figure \ref{fig:ablation}, the decoded 3D poses of $[\bm p_0, \bm p_1]$ starts with the posture of $\bm p_0$, while gradually putting down the leg and blending the arm towards the pose of $\bm p_1$.
Furthermore, for arbitrary combination of learned choreographic memory codes, the decoders can synthesize fluent movement based on the represented dance positions, which can be referred to the supplementary video.
With such characteristic, the choreography process becomes interpretable in proposed  {\textit{Bailando}} as a process of selecting and sorting the quantized dance positions from the learned choreographic memories, instead of a black box as most previous works. 
%
%

\vspace{-6pt}
\section{Discussion and Conclusion}
\vspace{-4pt}
\label{sec:conclusion}
In this paper, we address the spatial and temporal challenges of 3D dance generation  by proposing a novel framework named \textit{{Bailando}}, which is composed of a choreographic memory to address the spatial constraint by encoding and quantizing dancing-style poses, and an actor-critic GPT to realize the temporal coherency with music that translates and aligns various motion tempos and music beats. Experiments on the standard benchmark (\ie, AIST++ dataset) along with user studies show that \textit{{Bailando}} achieves state-of-the-art performance both qualitatively and quantitatively.


\small
{
{\noindent\textbf{Acknowledgement.}}
{This research is supported by the National Research Foundation, Singapore under its AI Singapore Programme (AISG Award No: AISG2-PhD-2022-01-031).
This research is conducted in collaboration with SenseTime.
This work is supported by NTU NAP and A*STAR through the Industry Alignment Fund - Industry Collaboration Projects Grant.
We thank Ruilong Li, Shan Yang and Zhiyuan Chen for their  help in this work.
}
}

{\small
\bibliographystyle{ieee_fullname}
\bibliography{egbib.bib}
}

\end{document}